\documentclass[twocolumn,pra,aps,superscriptaddress]{revtex4-2}
\usepackage[colorlinks=true, citecolor=blue, urlcolor=blue, linkcolor=red]{hyperref}
\usepackage{orcidlink,amsmath,amssymb}

\begin{document}
\title{Decoherence suppression in quantum dense coding}
\author{Xi Yan}
\affiliation{School of Physical Science and Technology \& Lanzhou Center for Theoretical Physics, Lanzhou University, Lanzhou 730000, China}
\affiliation{Key Laboratory of Quantum Theory and Applications of MoE \& Key Laboratory of Theoretical Physics of Gansu Province, Lanzhou University, Lanzhou 730000, China}
\author{Wen-Jie Zhang}
\affiliation{School of Physical Science and Technology \& Lanzhou Center for Theoretical Physics, Lanzhou University, Lanzhou 730000, China}
\affiliation{Key Laboratory of Quantum Theory and Applications of MoE \& Key Laboratory of Theoretical Physics of Gansu Province, Lanzhou University, Lanzhou 730000, China}
\author{Jun-Hong An\orcidlink{0000-0002-3475-0729}}
\email{anjhong@lzu.edu.cn}
\affiliation{School of Physical Science and Technology \& Lanzhou Center for Theoretical Physics, Lanzhou University, Lanzhou 730000, China}
\affiliation{Key Laboratory of Quantum Theory and Applications of MoE \& Key Laboratory of Theoretical Physics of Gansu Province, Lanzhou University, Lanzhou 730000, China}
	
\begin{abstract}
By harnessing entanglement to transmit information beyond the limits of classical communication, quantum dense coding serves as a key component in various quantum technologies. However, in the noisy intermediate-scale quantum era, noise-induced decoherence poses inevitable challenges to its practical implementation. By investigating the non-Markovian decoherence dynamics of the two categories of quantum dense coding protocols realized in discrete- and continuous-variable systems, we demonstrate a physical mechanism to suppress the decoherence in quantum dense coding. It is found that the formation of a bound state in the energy spectrum of the entire system consisting of each subsystem and its local environment of the quantum channel restores the quantum advantage of dense coding in the presence of decoherence. Refreshing our understanding on the noisy effect on quantum dense coding, our finding offers a guide toward the practical realization of noise-tolerant quantum dense coding.	
\end{abstract}
\maketitle
	
\section{INTRODUCTION}
Quantum dense coding aims to achieve higher channel capacity than classical communication by leveraging entanglement as quantum channel \cite{PhysRevLett.69.2881}. As a fundamental protocol in quantum communication, it serves as a critical building block for various large-scale quantum technologies. It plays an essential role in quantum networks \cite{RevModPhys.95.045006,PhysRevLett.93.210501,PhysRevA.92.052330,PhysRevA.106.052607,Zhang2014}, quantum computation \cite{PhysRevLett.81.5932}, and various quantum communication schemes including quantum teleportation \cite{Bouwmeester1997,PhysRevLett.70.1895,Pirandola2015,Hu2023}, quantum key distribution \cite{PhysRevA.65.032302,https://doi.org/10.1049/iet-ifs.2017.0375}, and secure quantum direct communication \cite{PhysRevApplied.17.064011}. Depending on the nature of the encoded information, quantum dense coding is typically divided into discrete- \cite{PhysRevLett.76.4656,https://doi.org/10.1002/qute.201900011,PhysRevLett.118.050501} and continuous-variable \cite{RevModPhys.77.513,RevModPhys.84.621,Li:18,PhysRevLett.88.047904} schemes. It has been realized on multiple physical platforms, including photonic qubits \cite{PhysRevLett.76.4656}, optical modes \cite{PhysRevLett.88.047904}, nuclear magnetic resonance \cite{PhysRevA.61.022307}, and atomic systems \cite{PhysRevLett.93.040505,PhysRevLett.118.220501,PhysRevA.71.012304,PhysRevA.71.034304,Li2020,Li2013,khedif_intrinsic_2022}. To further enhance communication capacity, multimode multiplexing \cite{PhysRevLett.127.093601} and high-dimensional encoding schemes \cite{PhysRevA.65.022304,PhysRevLett.90.167903,yang_bidirectional_2019} have been proposed. Meanwhile, two branches of quantum dense coding known as probabilistic dense coding \cite{PhysRevA.54.1869,PhysRevA.72.012329} and controlled dense coding \cite{PhysRevA.63.054301} have also been developed. Furthermore, extending the original point-to-point scenario to distributed dense coding \cite{PhysRevA.92.052330,PhysRevLett.96.060502,PhysRevA.100.052304} opens broader perspectives for applications in quantum networks.

However, in the noisy intermediate-scale quantum era \cite{Preskill2018quantumcomputingin,Chen2023}, the quantum advantage of dense coding is inevitably undermined by noise-induced decoherence. Mitigating the destructive influence of decoherence, therefore, is a major challenge. An ideal dense coding protocol requires a noiseless quantum channel based on a maximally entangled pure state. Unfortunately, in noisy environments, quantum advantage is often completely lost \cite{PhysRevA.85.052306,PhysRevA.90.022319,Shadman_2010}. Depending on whether it affects one or both parties involved in the protocol, noise can be classified into one-sided noise and two-sided noise \cite{PhysRevA.84.042309,PhysRevA.92.052330}. Several strategies have been proposed to suppress decoherence, including external field control \cite{aili_dense_2019,xu_mitigating_2025}, parameter tuning \cite{Zhao2019,Shadman_2010}, and non-unitary operations based on dissipation engineering \cite{chen_improving_2024}. Studies of quantum dense coding in noisy channels have extensively based on the Born-Markov approximation \cite{PhysRevA.85.052306,PhysRevA.90.022319}. Within the Born-Markov framework, the quantum advantage vanishes in both schemes \cite{PhysRevA.85.052306,PhysRevA.90.022319,Shadman_2010,article,PhysRevApplied.13.034029,PhysRevApplied.19.064015,PhysRevLett.126.250501}. In contrast, although non-Markovian noise can transiently revive quantum advantages \cite{bylicka_non-markovianity_2014,PhysRevLett.123.040402,PhysRevA.102.012217,PhysRevLett.124.140502,PhysRevLett.131.050801}, its vanishing fate in the long-time limit has never been changed \cite{PhysRevA.84.042309,e24081114,PhysRevA.93.032135,Liu_2016,PhysRevA.109.032616,Zhao2019,PhysRevA.82.032340,islam_optimizing_2021,AIACHE2024129320}. Two key issues naturally arise. First, is the destruction effect of the decoherence on quantum dense coding ostensible or fundamental in physics? Second, can its quantum superiority be persistently preserved in the exact non-Markovian decoherence dynamics?

\begin{figure}[tbp]
\centering
\includegraphics[width=.99\columnwidth]{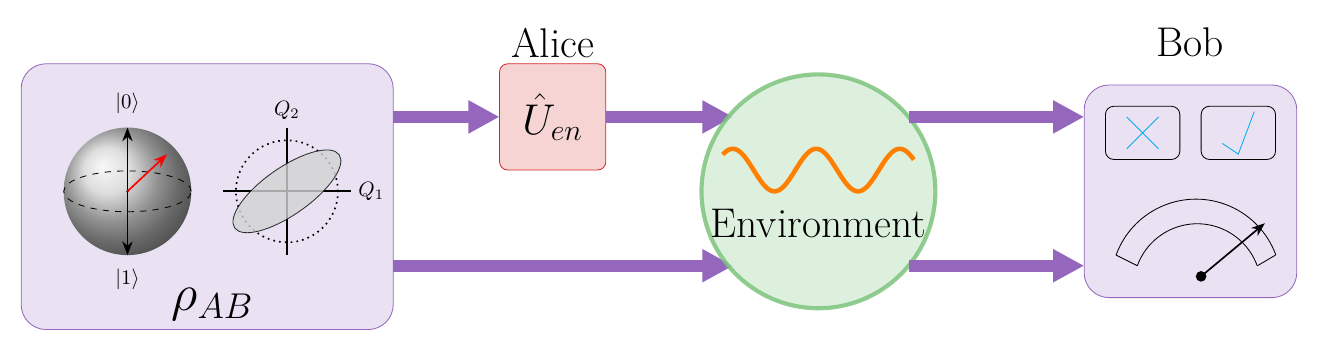}
\caption{Schematic diagrams of quantum dense coding. Encoding is realized by the unitary operations $\hat{U}_{en}=\hat{X}_j\in \{\hat{\sigma}_0, \hat{\sigma}_x, \hat{\sigma}_y, \hat{\sigma}_z\}$ and decoding is realized by Bell measurement for discrete-variable scheme. For continuous-variable scheme, encoding is via the displacement operations $\hat{U}_{en}=\hat{D}(\alpha_\text{in}/\sqrt{2})$ and the decoding is via the homodyne measurement.}\label{scheme}
\end{figure}
	
In this work, we address these issues by demonstrating that non-Markovian effects, when combined with bound state formation, can restore and sustain the quantum advantage of dense coding in both discrete- and continuous-variable systems.  By investigating the non-Markovian decoherence dynamics of both systems, we find that the performance of dense coding capacity is sensitively determined by the energy spectrum of the composite system formed by the channel subsystems and their environments. When an appropriate bound state emerges, both schemes achieve long-term capacities beyond their classical limits, yielding a sustained quantum advantage that outperforms previous results \cite{PhysRevLett.107.253602}. Our results thus identify the joint mechanism underlying this recovery as the interplay between non-Markovian effects and bound state formation. These findings deepen our understanding of robust quantum dense coding and offer a practical pathway toward mitigating decoherence in noisy environments.  
	
This paper is organized as follows. In Sec. \ref{Ideascm}, we introduce the ideal quantum dense coding protocols for discrete- and continuous-variable systems. Section \ref{noiseff} is devoted to analyzing their non-Markovian decoherence dynamics, emphasizing the crucial role of bound states formed between subsystems and their local environments. Numerical simulations validating our analytical findings are presented in Sec. \ref{numres}. We conclude the paper in Sec. \ref{discussion}. 

\section{Ideal quantum dense coding} \label{Ideascm}
As a fundamental quantum-information protocol, quantum dense coding uses entangled states to transfer information more efficiently than classical processes, as shown in Fig. \ref{scheme}. According to the properties of the Hilbert space of the systems used, it is categorized as discrete- \cite{Zhao2019} and continuous-variable schemes \cite{PhysRevD.95.105005}.
	
Discrete-variable quantum dense coding utilizes bipartite two-level systems (TLSs) to encode information \cite{PhysRevLett.69.2881}. Alice and Bob first share a pair of TLSs prepared in the maximally entangled state $|\Phi_{AB}^{+}\rangle =\frac{1}{\sqrt{2}}(|g_Ag_B\rangle +|e_Ae_B\rangle)$ \cite{PhysRevLett.69.2881,https://doi.org/10.1002/qute.201900011}. In order to transmit two bits of information to Bob, Alice encodes the information by applying one of the four Pauli operators, $\hat{X}_j\in \{\hat{\sigma}_0, \hat{\sigma}_x, \hat{\sigma}_y, \hat{\sigma}_z\}$, to her local system according to the information to be transmitted. While the identity operation $\hat{\sigma}_0$ retains the original $|\Phi_{AB}^{+}\rangle$, the application of $\hat{\sigma}_{x/y/z}$ yields the other three Bell states: $|\Psi_{AB}^{+}\rangle$, $|\Psi_{AB}^{-}\rangle$, $|\Phi_{AB}^{-}\rangle$, respectively. The four Bell states, each associated with a specific apriori probability $p_i$ \cite{PhysRevA.63.022302,Mirmasoudi_2018}. After receiving Alice's particle through the quantum channel, Bob performs a Bell-state measurement on the combined pair to decode the transmitted information. Because Alice's four local unitary operations generate four mutually orthogonal Bell states, Bob's measurement yields four distinct outcomes. Thus, in units of bits, quantum dense coding enables two separated parties to transfer two bits of classical information by transmitting only one TLS \cite{PhysRevResearch.2.023023}. By contrast, without entanglement, a single transmitted TLS admits only two distinguishable outcomes and thus carries at most one bit of information. The performance of such a protocol is characterized by the channel capacity, defined as the maximum of the Holevo quantity \cite{Shadman_2010,Zhao2019,PhysRevA.54.1869,PhysRevA.63.022302} 
\begin{equation}
C  = \max_{\{p_j, \rho_j\}} \big[ S(\bar{\rho}) - \sum_{j} p_j S(\rho_j) \big],\label{dv}
\end{equation}
$\rho_j=(\hat{X}_j\otimes I)\rho(\hat{X}_j\otimes I)$, with $\rho=|\Phi_{AB}^+\rangle\langle\Phi_{AB}^+|$, and $S(\rho)=-\mathrm{Tr}(\rho\log_2\rho)$ is the von Neumann entropy, $\overline{\rho}=\sum_{j}p_j\rho_j$ is the average state of the signal ensemble. For an optimal ensemble, Eq. \eqref{dv} is maximized at $p_j=1/4$ \cite{PhysRevA.63.022302}, yielding $C=1+S(\rho_B)-S(\rho)$ \cite{PhysRevA.84.042309,PhysRevResearch.2.023023}, where $\rho_B=\text{Tr}_A(\rho)$. In the ideal scenario of unitary dynamics governed by the Hamiltonian $\hat{H}_0=\omega_0\sum_{l=A}^{B}\hat{\sigma}^{\dagger}_l\hat{\sigma}_l$, with $\hat{\sigma}_l=|g_l\rangle\langle e_l|$, the shared Bell state evolves to $|\Phi_{AB}(t)\rangle=(e^{-2i\omega_0 t}|e_Ae_B\rangle+|g_Ag_B\rangle)/\sqrt{2}$. The four encoded states remain pure and mutually orthogonal, so that $ S[\rho_j(t)]=0$, while $\overline{\rho}(t)$ is maximally mixed with $S[\bar{\rho}(t)]=2$. Hence, we have $C(t)=2$, which surpasses the classical channel capacity $C_\text{C}=1$. 

Continuous-variable quantum dense coding is a protocol in which Alice and Bob share two entangled light beams prepared in a two-mode squeezed vacuum state $|\Phi_{AB}\rangle=\exp[ r(\hat{a}^{\dagger}\hat{b}^{\dagger}-\hat{a}\hat{b})]|0_A0_B\rangle $ \cite{PhysRevA.102.052425}. Alice encodes her classical messages $x_{\text{in}}$ and $p_{\text{in}}$ by applying a displacement operator $\hat{D}(\alpha_{\text{in}}/\sqrt{2})=\exp[(\alpha_{\text{in}}\hat{a}^{\dagger}-\alpha_{\text{in}}^*\hat{a})/\sqrt{2}]$ on her beam, where $\alpha_{\text{in}}=x_\text{in}+ip_\text{in}$ obeys a Gaussian probability distribution $P_{\alpha_{\text{in}}}=\exp[-|\alpha_\text{in}|^2/(2\sigma^2)]/(2\pi \sigma^2)$, with $\sigma$ being the standard deviation. The encoded beam is then transmitted to Bob. Bob measures the two quadrature operators $\hat{x}=\hat{x}_A-\hat{x}_B$ and $\hat{p}=\hat{p}_A+\hat{p}_B$ on the state of the two beams by homodyne measurement, where $\hat{x}_{j}=(\hat{a}_j+\hat{a}^{\dagger}_j)/{\sqrt{2}}$ and $\hat{p}_{j}=(\hat{a}_j-\hat{a}^{\dagger}_j)/(\sqrt{2}i)$ for $j=A,B$. The conditional probability that the measured result is $x$ and $p$ is denoted by $P(\alpha| \alpha_{\text{in}})$, with $\alpha=x+ip$. Thus, we have the unconditional output probability distribution as $P(\alpha)=\int {\rm d}^2\alpha_{\text{in}}P_{\alpha_\text{in}}P(\alpha|\alpha_\text{in})$. The channel capacity of continuous-variable dense coding is defined as the maximum mutual information \cite{PhysRevLett.70.363}
\begin{equation}
I=\int {\rm d}^2\alpha {\rm d}^2\alpha_{\text{in}}P(\alpha|\alpha_\text{in})P_{\alpha_\text{in}}\ln\frac{P(\alpha|\alpha_\text{in})}{P(\alpha)}.\label{mtif}
\end{equation}
Under the unitary dynamics governed by the Hamiltonian $\hat{H}_0=\sum_{l=A}^B\omega_0\hat{a}^{\dagger}_{l}\hat{a}_{l}$ of the two light beams, the state evolves to
\begin{equation}
|\Phi_{AB}(t)\rangle=\cosh^{-1}r\exp[e^{-2i\omega_0 t} \tanh r \hat{a}^{\dagger}\hat{b}^{\dagger}]|0_A0_B\rangle.
\end{equation}
The common eigenstate of $\hat{x}$ and $\hat{p}$ is $|\Phi_{x,p}\rangle=\frac{1}{\sqrt{2\pi}}\exp(-{|\alpha|^2\over4}-i{px\over 2}+{\alpha\hat{a}^{\dagger}-\alpha^{*}\hat{b}^{\dagger}\over\sqrt{2}}+\hat{a}^{\dagger}\hat{b}^{\dagger})$. Therefore, the conditional probability of obtained $\alpha$ is calculated to be 
\begin{eqnarray}
P(\alpha|\alpha_\text{in})&\equiv&|\langle\Phi_{x,p}|\hat{D}(\alpha_\text{in}/\sqrt{2})|\Phi_{AB}(t)\rangle|^2\nonumber\\
&=&\exp[-|\alpha-\alpha_{\text{in}}|^2/(2{S_t})]/(2\pi S_t), 
\end{eqnarray} where $S_t=\cosh2r-\sinh2r\cos2\omega_0t$. Performing the Gaussian integration of $\alpha_\text{in}$ under the probability distribution $P_{\alpha_\text{in}}$, we obtain the unconditional output probability $P(\alpha)=\exp[{-|\alpha|^2\over 2(S_t+\sigma^2)}]/[2\pi (S_t+\sigma^2)]$. Substituting $P(\alpha)$ into Eq. \eqref{mtif}, we can derive the mutual information as $I(t)=\ln(1+\sigma^2/S_t)$. At $t=n\pi/\omega_0$, with $n\in\mathbb{Z}$, $I(t)$ reaches its maximum $I_\text{max}=\ln(1+e^{2r}\sigma^2)$. Note that $I_\text{max}$ diverges in the limit $\sigma\to\infty$. Being valid for any squeezing parameter $r$, this reflects the transmission of an unbounded amount of classical information. However, such a scenario would require infinite energy for encoding. Thus, a finite-energy constraint \cite{PhysRevA.90.022301} is typically imposed to allow fair comparisons between different schemes. For the two-mode squeezed vacuum state, the average photon number transmitted through the channel is $\bar{n}=\sigma^2+\sinh^2r$ for a given squeezing parameter $r$. Constrained by $\bar{n}$, one can optimize $\sigma$ to further maximize $I_\text{max}$ to obtain the channel capacity as \cite{PhysRevA.90.022301}
\begin{equation}
C=\ln(1+\bar{n}_\text{opt}+\bar{n}_\text{opt}^2),\label{cvevo} 
\end{equation}with $\sigma_\text{opt}^2=\sinh(2r)/2$ and $\bar{n}_\text{opt}=e^r\sinh r$. As a classical benchmark, a single-mode coherent-state transmitter with mean photon number $\bar{n}_\text{coh}=|\alpha|^2$ yields \cite{RevModPhys.58.1001,PhysRevLett.96.080502}
\begin{equation}
 C_\mathrm{coh}=\ln(1+\bar{n}_\text{coh}).   \label{coh}
\end{equation}
The dense coding based on the two-mode squeezed vacuum state always outperforms the one based on the coherent state under the equal constraint $\bar{n}_\text{coh}=\bar{n}_\text{opt}$.
	
\section{Noisy effects} \label{noiseff}
The quantum channel inevitably suffers from noise-induced decoherence, leading to detrimental consequences to quantum dense coding. Decoherence is broadly classified into dissipation and dephasing, depending on whether energy exchange between the system and the environment occurs or not. A microscopic description of these decoherence processes is provided by the theory of open quantum system. Starting from the unitary dynamics of the total system consisting of the quantum system and its environment and tracing out the environmental degrees of freedom, a reduced dynamics of the open system is achieved. In this process, the Born-Markov approximation is commonly employed. The primary decoherence sources of quantum dense coding are spontaneous emission from the two TLSs and photon dissipation in the quantum channels \cite{PhysRevLett.76.4656,PhysRevLett.76.4656,https://doi.org/10.1002/qute.201900011}. Conventionally, phenomenological descriptions have been achieved via the Kraus-operator representation in the case of discrete-variable channels \cite{PhysRevA.92.052330} and through the introduction of an imperfect beam splitter for continuous-variable channels \cite{yang_optimal_2015}. They are equivalent to the Born-Markov approximation. In this work, we go beyond this approximation \cite{PhysRevLett.101.150402,Liu_2016} by exactly evaluating the effects of local dissipative noises on both types of quantum dense coding. 

Consider that each subsystem of the quantum channel interacts with a dissipative environment. The Hamiltonians for both discrete- and continuous-variable quantum channels are written in a unified manner as $\hat{H}=\sum_{l=A}^{B}\hat{H}_l$, with
\begin{equation}
\hat{H}_l=\omega_0\hat{o}_l^\dagger\hat{o}_l+\sum_{k}[\omega_{k}\hat{b}_{l,k}^{\dagger}\hat{b}_{l,k}+ (g_k\hat{b}_{l,k}^{\dagger}\hat{o}_l+ \text{h.c.})],\label{Hamiltonian}
\end{equation}
where $\hat{b}_{l,k}$ is the annihilation operator for the $k$th mode of the environment with frequency $\omega_k$ coupled to the $l$th subsystem of the quantum channel, $g_k$ is the coupling strength, and $\hat{o}_l=\hat{\sigma}_l$ for the discrete-variable case and $\hat{o}_l=\hat{a}_l$ for the continuous-variable case are the annihilation operators of the two subsystems. The coupling strength is further characterized by the spectral density $J(\omega)=\sum_k|g_k|^2\delta(\omega-\omega_k)$, which is generally modeled as $J(\omega)=\eta\omega^s\omega_c^{1-s}e^{-\omega/\omega_c}$, where $\eta$ is a dimensionless coupling constant, $\omega_c$ is a cutoff frequency, and $s$ is the Ohmicity parameter. The environment is classified into sub-Ohmic for $0 < s < 1$, Ohmic for $s = 1$, and super-Ohmic for $s > 1$ \cite{RevModPhys.59.1}. When the environments are initially in the vacuum state, their degrees of freedom for both of the two cases can be traced exactly. The dynamics of the two types of quantum channels are governed by the non-Markovian master equation \cite{Ji2026}
\begin{equation}
\dot{\rho}(t)=\sum_{l=1,2}{-i\Omega(t)[\hat{o}_l^{\dagger}\hat{o}_l,\rho(t)]+\Gamma(t)\check{\mathcal{L}}_l\rho(t)},
\label{rho}
\end{equation}where $\rho(t)$ is the reduced density matrix of the quantum channel formed by the two subsystems, $\check{\mathcal{L}}_l\cdot=2\hat{o}_l\cdot\hat{o}^\dagger_l-\{\hat{o}^\dagger_l\hat{o}_l,\cdot\}$ is the Lindblad superoperator, $\Gamma(t)\equiv-\text{Re}[\dot{u}(t)/u(t)]$ is the decay rate, and $\Omega(t)\equiv-\text{Im}[\dot{u}(t)/u(t)]$ is the renormalized frequency. Acting as a decoherence factor, the function $u(t)$ satisfies the integro-differential equation
\begin{equation}
\dot{u}(t)+\mathit{i}\omega_0 u(t)+\int_0^t {\rm d}\tau f(t-t')u(t')=0,\label{uu}
\end{equation}
under the initial condition $u(0)=1$. Here, $f(x)=\int_0^\infty {\rm d}\omega J(\omega)e^{-\mathit{i}\omega x}$ is the environmental correlation function. The convolution in Eq. \eqref{uu} renders the dynamics non-Markovian. Under special cases in which the system-environment interaction is weak and the characteristic time scale of the environmental correlation function is much shorter than that of the system, the Born-Markov approximation is applied by replacing $u(\tau)$ with $u(t)$ and extending the upper limit of integration from $t$ to infinity. The resulting solution is $u_{\text{BMA}}(t)\backsimeq e^{-[\kappa+\mathit{i}(\omega_0+\Delta_{\omega_0})]t} $, where $\kappa=\pi J(\omega_0)$ and $\Delta_{\omega_0}=\mathcal{P}\int_0^\infty\frac{J(\omega)}{\omega_0-\omega}{\rm d}\omega $, with $\mathcal{P}$ being the principal Cauchy value.
	
In the general non-Markovian case, the solution of Eq. \eqref{uu} must resort to a numerical calculation. However, its long-time form is analytically solvable, which allows us to analyze the asymptotical behavior of $C(t)$. Applying the Laplace transform to Eq. \eqref{uu}, we obtain $\tilde{u}(p)=[p+\mathit{i}\omega_0+\int_0^\infty\frac{J(\omega)}{p+\mathit{i}\omega}{\rm d}\omega]^{-1}$. The function $u(t)$ is obtained via the inverse Laplace transform to $\tilde{u}(p)$, which involves identifying the poles of $\tilde{u}(p)$ through
\begin{equation}
Y(E)\equiv\omega_0-\int_0^{\infty}\frac{J(\omega)}{\omega-E}{\rm d}\omega=E,\quad (E=ip).\label{eb}
\end{equation}
It is interesting to find that the roots of Eq. \eqref{eb} correspond exactly to the eigenenergies of each subsystem and its local environment within the single-excitation subspace. To demonstrate this, we expand the eigenstate of $\hat{H}_l$ as $|\Phi\rangle=(x\hat{o}_l^{\dagger}+\sum_ky_k\hat{b}_k^{\dagger})|\emptyset_l, {0_k}\rangle$, where $|\emptyset_l\rangle=|g\rangle$ for the discrete-variable case and $|\emptyset_l\rangle=|0\rangle$ for the continuous-variable case. Substituting this form into $\hat{H}_l|\Phi\rangle=E|\Phi\rangle$ produces exactly Eq. \eqref{eb} in the continuous limit of the environmental frequencies. This result reveals that the decoherence dynamics of both types of quantum channels is essentially governed by the energy-spectrum characteristics of each local subsystem and its environment. This finding provides an insightful guideline for controlling the decoherence of the open quantum system by engineering the energy spectrum structure of the total system formed by the system and its environment. In particular, Eq. \eqref{eb} has exactly one isolated root $E_b$ in the regime $E < 0$ provided $Y(0)<0$, because $Y(E)$ is a monotonically decreasing function in this regime. The corresponding eigenstate is referred to as the bound state. In contrast, for $E> 0$, the integral becomes divergent, and Eq. \eqref{eb} has an infinite number of roots, which form a continuous energy band. Subsequently, applying the inverse Laplace transform via the contour integration and the residue theorem leads to
\begin{equation}
\hspace{-0.2cm}u(t)=Ze^{-iE_bt}+\int_0^\infty\frac{J(E)e^{-iEt}{\rm d}E}{(E-\omega_0-\Delta_E)^2+[\pi J(E)]^2},\label{unm}
\end{equation}where the first term with $Z=[1+\int_0^\infty\frac{J(\omega)d\omega}{(E_b-\omega)^2}]^{-1}$ represents the contribution of the bound state and the second term arises from the branch cut formed by the band energies. In the long-time regime, the second term tends toward zero due to out-of-phase interference of the continuously varying oscillation frequencies $E$. Thus, if no bound state is present, then $u(\infty)=0$, which indicates a complete decoherence. In contrast, if a bound state exists, then $u(\infty)\simeq Ze^{-iE_b t}$, which characterizes a decoherence suppression. From the condition $Y(0)<0$, we can evaluate that the bound state is present when $\omega_0  \leq \eta\omega_c\gamma(s)$, where $\gamma(s)$ denotes the Euler's gamma function, for the Ohmic-family spectral density. Therefore, we can effectively suppress the decoherence of the quantum channels by properly manipulating the system frequency $\omega_0$ and the parameters in the spectral density to support the formation of the bound state. To validate this prediction, we proceed to explicitly investigate the effect of dissipative decoherence on the two types of quantum dense coding protocols.

\subsection{Discrete-variable system}
Solving Eq. \eqref{rho} with $\hat{o}=\hat{\sigma}$ for the initial state $\rho(0) = |\Phi_{AB}^{+}\rangle\langle\Phi_{AB}^{+}|$, we obtain
\begin{eqnarray}\label{11}
\rho(t) &=&\{[P_t|e\rangle\langle e|+(1-P_t)|g\rangle\langle g|]^{\otimes2} + |g\rangle\langle g|^{\otimes2} \notag \\
&&+[u^2(t)|e\rangle\langle g|^{\otimes2}+\text{h.c.}]\}/2,\label{noisydv}
\end{eqnarray}
where $P_t=|u(t)|^2$. Then the average
state of the signal ensemble becomes $\bar{\rho}(t)={P_t\over 4}(|eg\rangle\langle eg|+|ee\rangle\langle ee|)+{2-P_t\over 4}(|gg\rangle\langle gg|+|ge\rangle\langle ge|)$. The channel capacity is calculated to be
\begin{equation}
C(t)={1\over\ln4}\Big[(h-f)\ln\frac{h-f}{2}+(h+f)\ln\frac{h+f}{2}+g\Big],\label{dvnoise}
\end{equation}
where  $f=\sqrt{1-4x}$, $g=4(x\ln x-y\ln y-z\ln z)$, $h=1-2x$, $x=(P_t-P_t^2)/2$, $y=(2-P_t)/4$, and $z=P_t/4$. It is easily checked that $u_\text{ideal}(t)=e^{-i\omega_0t}$ and thus $P_t=1$ and $C_\text{ideal}(t)=2$ in the noiseless limit, recovering the ideal result. Under the Born-Markov approximate decoherence, $P_t^\text{BMA}=e^{-2\kappa t}$ decays to zero, so that $C_\mathrm{BMA}(\infty)=1$ in the long-time limit. It indicates that the quantum advantage of dense coding is completely destroyed by the Markovian dissipative noise.

In the non-Markovian case, it is natural to expect that $C(t)$ tends to the classical limit with time in the absence of the bound state. We therefore focus on the case where the bound state is present. Inserting $\lim_{t\to\infty}P_t=Z^2$ into Eq. \eqref{dvnoise} gives
\begin{equation}
C(\infty)={1\over\ln4}\Big(f'\ln\frac{f'}{2}+h'\ln\frac{h'}{2}+g'\Big),\label{dvz}
\end{equation} 
where $f'=1-x'-\sqrt{1-2x'}$, $h'=1-x'+\sqrt{1-2x'}$, $g'=4({x'\over 2}\ln {x'\over 2}-y'\ln y'-z'\ln z')$, $x'=Z^2-Z^4$, $y'=(2-Z^2)/4$ and $z'=Z^2/4$. It surpasses the classical channel capacity $C_\text{C}=1$ as long as $P_t > 0.76$, or equivalently $Z > 0.87$. Therefore, thanks to the formation of the bound state, the quantum superiority of dense coding is persistently preserved in the non-Markovian dynamics. 

\subsection{Continuous-variable system}
Within the path-integral influence functional framework, the solution of Eq. \eqref{rho} with $\hat{o}=\hat{a}$ is written as
\begin{eqnarray}
\rho (\pmb{\bar{\alpha}}_{f},\pmb{\alpha }_{f}^{\prime };t)&=&\int {\rm d}\alpha (\pmb{\alpha }_{i}){\rm d}\alpha (\pmb{\alpha }_{i}^{\prime })\mathcal{J}(\pmb{\bar{\alpha}}_{f},\pmb{\alpha }_{f}^{\prime};t|\pmb{\alpha}_{i},\bar{\pmb{\alpha }}_{i}^{\prime };0)  \notag   \\
&&\times \rho (\pmb{\bar{\alpha}}_{i},\pmb{\alpha }_{i}^{\prime };0),\label{rout}
\end{eqnarray}
where $\rho(\pmb{\bar{\alpha}}_f,\pmb{\alpha}_{f}^{\prime};t) =\langle\bar{\pmb{\alpha}}_f|\rho(t)|\pmb{\alpha}_f'\rangle$ is the reduced density matrix $\rho(t)$ in the coherent-state representation, and $\mathcal{J}(\pmb{\alpha}_f,\pmb{\bar{\alpha}}_f^{\prime};t|\pmb{\alpha}_i,\pmb{\bar{\alpha}}_i;0)$ is the effective propagator. Here we adopt the coherent-state representation $|\pmb{\alpha}\rangle=\prod_{l=A,B}|\alpha_l\rangle$, with $|\alpha_l\rangle=\exp(\alpha_l\hat{\alpha}_l^{\dagger})|0_l\rangle$, which satisfies the resolution of identity $\int {\rm d}\mu(\pmb{\alpha})|\pmb{\alpha}\rangle\langle \bar{\pmb{\alpha}}|=1$ with the integration measures $ {\rm d}\mu(\pmb{\alpha})\equiv\prod_{l}e^{-\bar{\alpha}_l\alpha_l}\frac{{\rm d}^2\alpha_l}{\pi}$, and $\bar{\pmb{\alpha}}$ is the complex conjugate of $\pmb{\alpha}$. The effective propagator reads
\begin{eqnarray}
&&\mathcal{J}(\pmb{\bar{\alpha}}_{f},\pmb{\alpha }_{f}^{\prime};t|\pmb{\alpha}_{i},\bar{\pmb{\alpha }}_{i}^{\prime };0)=\exp\big\{\sum_{l=1,2}\big[u(t)\bar{\alpha}_{lf}\alpha _{li}  \notag \\
&&~~~~~+\bar{u}(t)\bar{\alpha}_{li}^{\prime }\alpha _{lf}^{\prime}+[1-\left\vert u(t)\right\vert ^{2}]\bar{\alpha}_{li}^{\prime }\alpha_{li}\big]\big\}.  \label{pro}
\end{eqnarray}
The initial state being the two-mode squeezed state takes the form 
$\rho (\boldsymbol{\bar{\alpha}}_i,\boldsymbol{\alpha }^{\prime }_i;0)=\cosh ^{-2}r e^{-\tanh  r (\bar{\alpha}_{1}\bar{\alpha}_{2}+\alpha _{1}^{\prime }\alpha _{2}^{\prime })}$ in the coherent-state representation. Substituting it and Eq. \eqref{pro} into Eq. \eqref{rout} and evaluating the Gaussian integral, we obtain the exact evolution state as
\begin{eqnarray}
\rho(\pmb{\bar{\alpha}}_f,\pmb{\alpha}_{f}^{\prime};t) 
&\equiv& \mathcal{A} \exp[ \mathcal{B}\bar{\alpha}_{1f}\bar{\alpha}_{2f} + \bar{\mathcal B}\alpha_{1f}^{\prime}\alpha_{2f}^{\prime} \nonumber \\
&& +\mathcal{C}(\bar{\alpha}_{1f}\alpha_{1f}^{\prime}+\bar{\alpha}_{2f}\alpha_{2f}^{\prime}) ], \label{eq:first}
\end{eqnarray}
with $\mathcal{A}=x\cosh^{-2}r$, $\mathcal{B}=x u(t)^2\tanh r$, $\mathcal{C}=x|u(t)|^2(1-|u(t)|^2)\tanh^2r$, and $x=[1-(1-|u(t)|^2)^2\tanh^2r]^{-1}$. The density matrix then reads $\rho(t)=\int {\rm d}\mu({\pmb\alpha}_f){\rm d}\mu({\pmb\alpha}_f')\rho(\pmb{\bar{\alpha}}_f,\pmb{\alpha}_{f}^{\prime};t) |{\pmb\alpha}_f\rangle\langle \bar{\pmb\alpha}_f'|$. After Alice's encoding operation, the state becomes $\tilde{\rho}(t)=\hat{D}(\alpha_{\text{in}}/\sqrt{2})\rho(t)\hat{D}^{\dagger}(\alpha_{\text{in}}/\sqrt{2})$. It is straightforward to calculate the conditional probability for Bob's measurement outcome $\alpha=x+ip$ to be
\begin{equation}
    P(\alpha|\alpha_{\text{in}})=\exp[-|\alpha-\alpha_{\text{in}}|^2/(2F_t)]/(2\pi F_t),
\end{equation}with $F_t=(|1-\mathcal{B}|^2-\mathcal{C}^2)/\mathcal{A}$, and the unconditional probability to be $P(\alpha)=\exp[-{|\alpha|^2\over2(F_t+\sigma^2)}]/[2\pi(F_t+\sigma^2)]$. Then, the mutual information in Eq. \eqref{mtif} is 
\begin{equation}
    I(t)=\ln(1+\sigma^2/F_t).\label{mutinf}
\end{equation}
In the noiseless limit, $\mathcal{A}_\text{ideal}=\cosh^{-2}r$, $\mathcal{B}_\text{ideal}=e^{-2i\omega_0 t}\tanh r$, and $\mathcal{C}_\text{ideal}=0$, which reduce $F_t$ to $S_t$. Under the Born-Markov approximation, $F_t^\text{BMA}=1+e^{-2\kappa t}[\tanh r-\cos(2\omega_0t)]\sinh(2 r)$, where the frequency shift $\Delta_{\omega_0}$ has been renormalized into the bare frequency $\omega_0$. At $t=n\pi/\omega_0$, $I(t)$ reaches its local maximum $I_\text{max}(t)=\ln[1+\sigma^2/(1+e^{-2\kappa t}(e^{-2r}-1))]$. Under the same optimization condition as the ideal case in Eq. \eqref{cvevo}, we obtain the channel capacity as 
\begin{equation}
C_\text{BMA}(t)=\ln\Big[1+{\bar{n}_\text{opt}(1+\bar{n}_\text{opt})\over 1+2\bar{n}_\text{opt}(1-e^{-2\kappa t})}\Big].
\end{equation} In the lon-time limit, $C_\text{BMA}(\infty)=\ln\big[1+{\bar{n}_\text{opt}(1+\bar{n}_\text{opt})\over 1+2\bar{n}_\text{opt}}\big]$, which is even worse than $C_\text{coh}$. Thus, the quantum superiority of the continuous-variable dense coding is completely destroyed by the Born-Markovian decoherence \cite{RevModPhys.58.1001}. 
 
In the general non-Markovian dynamics, we focus on the case in the presence of the bound state. The substitution of the long-time $\lim_{t\to\infty}u(t)=Ze^{-iE_b t}$ into Eq. \eqref{mutinf} results in
\begin{equation}
\lim_{t\to\infty} I(t)=\ln\big[1+{\sigma^2\over1+Z^2\sinh(2r)(\tanh r-\cos(2E_bt))}\big].
\end{equation}
Its maximum value $I_\text{max}=\ln[1+\sigma^2/(1+Z^2 (e^{-2r}-1))]$ is achieved when $t=n\pi/E_b$. Under the same optimization condition as the ideal case in Eq. \eqref{cvevo}, we obtain the channel capacity as
\begin{equation}
C(\infty)=\ln\Big[1+{\bar{n}_\text{opt}(1+\bar{n}_\text{opt})\over 1+2\bar{n}_\text{opt}(1-Z^2)}\Big].\label{cvz}
\end{equation}This value always exceeds the Born-Markov approximate result $C_\mathrm{BMA}(\infty)$. A comparison of Eq. \eqref{coh} and \eqref{cvz} further shows that it exceeds the channel capacity of the coherent state $C_\text{coh}$ as long as $Z>1/\sqrt{2}$. Therefore, the quantum superiority is retrieved in the non-Markovian dynamics due to the formation of the bound state. 

\begin{figure}
\includegraphics[width=1.0\columnwidth]{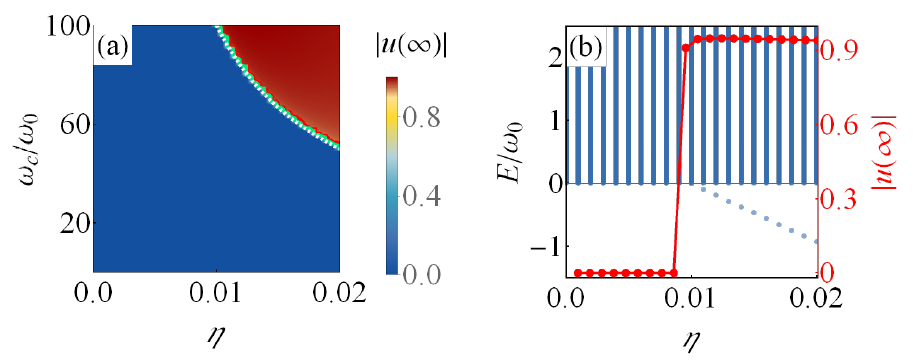}\\
\includegraphics[width=1.0\columnwidth]{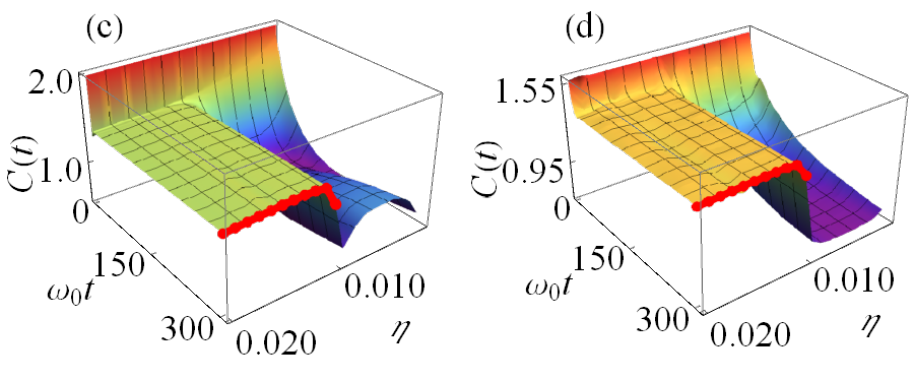}\\
\includegraphics[width=1.0\columnwidth]{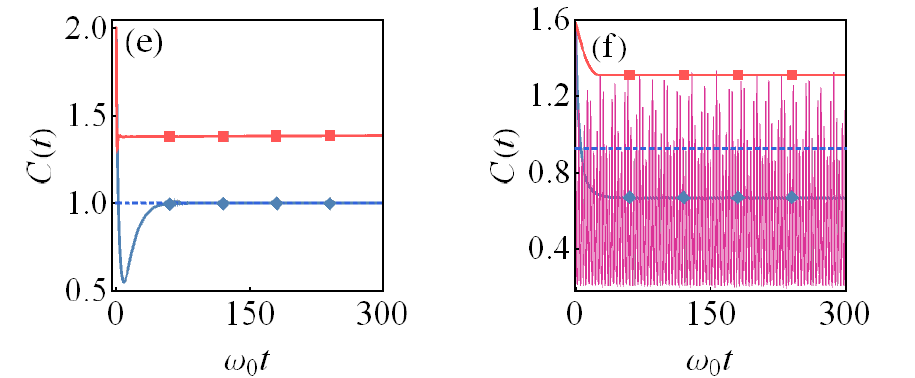}
\caption{(a) Decoherence factor $|u(t)|$ at $t=300\omega_0^{-1}$ in the $\eta$-$\omega_c$ plane via numerically solving Eq. \eqref{unm} at $s=1.0$.The red, green and white dashed curves denote $Z=0.87$, $1/\sqrt{2}$, and the bound state thresholdy, respectively.  (b) Energy spectrum (blue dots) obtained from Eq. \eqref{eb} and $Z$ (red lines) as a function of $\eta$ when $\omega_c=100\omega_0$. Evolution of the channel capacity $C(t)$ in different $\eta$ for (c) the discrete-variable system calculated via Eq. \eqref{dvz}. (d) Evolution of the envelope formed by the local maxima of $C(t)$ for the continuous-variable system calculated via Eq. \eqref{cvz}. The red curve shows analytical results evaluated from the bound-state energy $E_b$. (e) and (f) plot the non-Markovian (magenta curves),  Born-Markov approximate (blue solid lines), and classical (blue dashed lines) dynamics of $C(t)$ for the discrete- and continuous-variable systems.} \label{eta}
\end{figure}	

\begin{figure}[tbp]
\centering
\includegraphics[width=1.0\columnwidth]{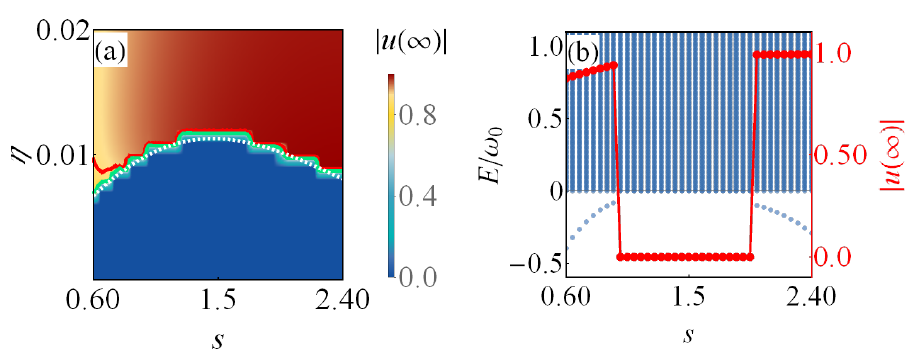}\\
\includegraphics[width=1.0\columnwidth]{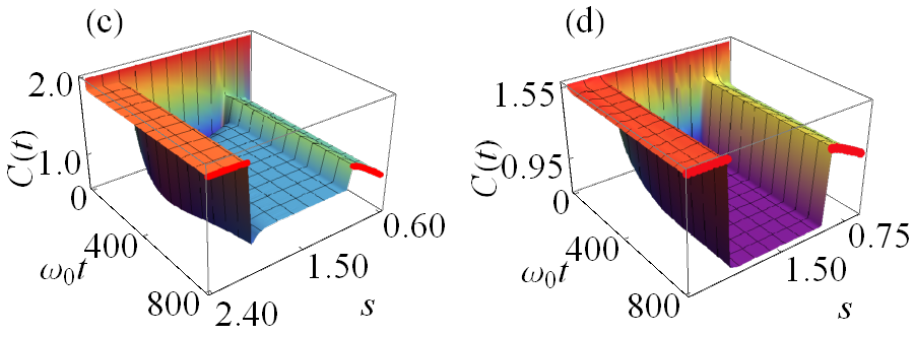}\\
\includegraphics[width=1.0\columnwidth]{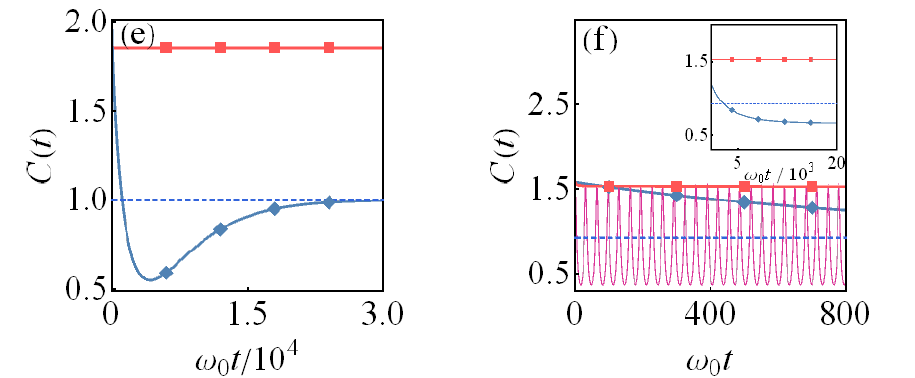}
\caption{(a) $|u(t)|$ at $t=800\omega_0^{-1}$ in the $s$-$\eta$ plane at $\omega_c/\omega_0=100$. The red, green and white dashed curves denote $Z=0.87$, $1/\sqrt{2}$, and the bound state thresholdy, respectively. (b) Energy spectrum (blue dots) and $Z$ (red lines) versus $s$ at $\eta=0.01$. (d) Evolution of the capacity $C(t)$ (c) and the envolope of the local maxima of $C(t)$ (d) versus $\omega_c/\omega_0$ for discrete- and continuous-variable systems. The red curves denote the analytical results.  (e) and (f) plot the non-Markovian (magenta curves),  Born-Markov approximate (blue solid lines), and classical (blue dashed lines) dynamics of $C(t)$ for the discrete- and continuous-variable systems.} \label{s}
\end{figure}	

\begin{figure}[tbp]
\centering
\includegraphics[width=.99\columnwidth]{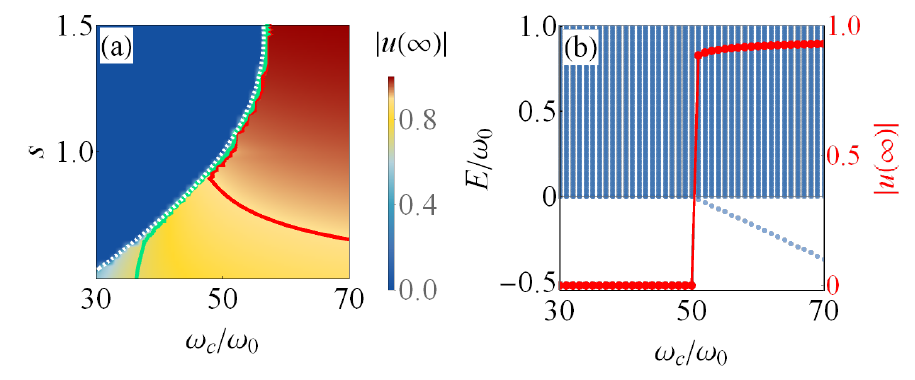}\\
\includegraphics[width=.99\columnwidth]{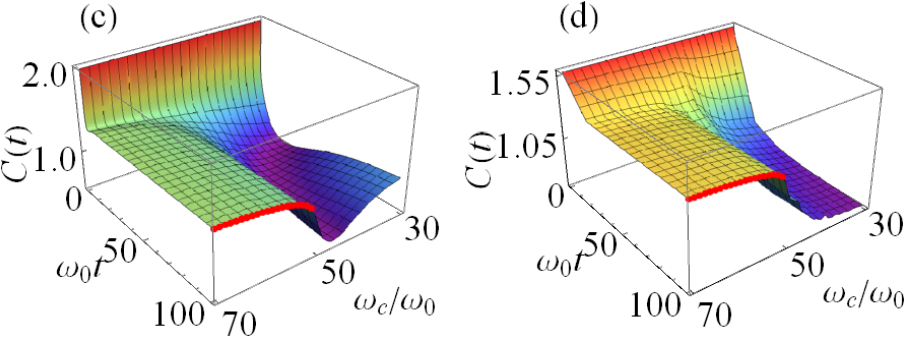}\\
\includegraphics[width=1.0\columnwidth]{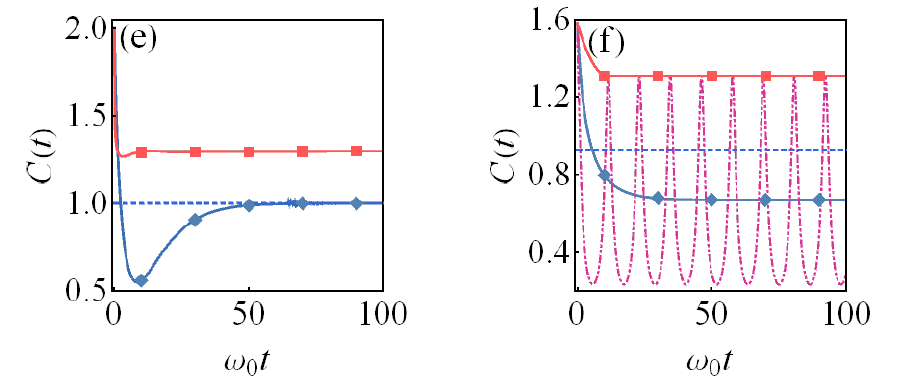}
\caption{(a) $|u(t)|$ at $t=100\omega_0^{-1}$ in the $\omega_c$-$s$ plane at $\eta=0.02$. The red, green and white dashed curves denote $Z=0.87$, $1/\sqrt{2}$, and the bound state thresholdy, respectively. (b) Energy spectrum (blue dots) and $Z$ (red lines) versus $\omega_c/\omega_0$ at $s=1.0$. (d) Evolution of the capacity $C(t)$ (c) and the envolope of the local maxima of $C(t)$ (d) versus $\omega_c/\omega_0$ for discrete- and continuous-variable systems. The red curves denote the analytical results. (e) and (f) plot the non-Markovian (magenta curves),  Born-Markov approximate (blue solid lines), and classical (blue dashed lines) dynamics of $C(t)$ for the discrete- and continuous-variable systems.} \label{n}
\end{figure}	

\section{Numerical results}\label{numres}
To systematically reveal how the bound state affects the quantum dense coding, we perform numerical simulations for both discrete- and continuous-variable systems. Figure \ref{eta}(a) illustrates the long-time limit of $|u(t)|$ as a function of the coupling strength $\eta$ and the cutoff frequency $\omega_c$. The bound state is present above the white dashed line. The red curve marks $Z = 0.87$, above which Eq. \eqref{dvz} predicts a quantum advantage for the discrete-variable system. The green curve marks $Z = 1/\sqrt{2}$, above which Eq. \eqref{cvz} predicts the same for the continuous-variable system. Figure \ref{eta}(b) plots the energy spectrum of $\hat{H}_l$ versus $\eta$ at $\omega_c/\omega_0=100$. A bound state emerges at $\eta \geq 0.01$, consistent with the analytical criterion $\eta\geq\omega_0/[\gamma(s)\omega_c]$. Once the bound state is formed, $|u(t)|$ evolves towards a finite asymptotic value, matching $Z$ derived from the bound-state eigenenergy $E_b$. The resulting channel capacity $C(t)$ is shown in Fig. \ref{eta}(c) for the discrete-variable system and \ref{eta}(d) for the continuous-variable system. An obvious change occurs at the bound-state formation threshold. When a bound state is present, $C(t)$ approaches a finite value for both systems, whereas its absence leads to irreversible decay. Figures \ref{eta}(e) and \ref{eta}(f) compare the exact non-Markovian, the Born-Markov approximate, and the classical results for both systems at $\eta=0.02$. In the discrete-variable system, the non-Markovian capacity outperforms both the Born-Markovian and the classical limits. For the continuous-variable system, although $C(t)$ exhibits sustained periodic oscillations, the profile formed by its local maxima exceeds both the Born-Markov result and $C_\text{coh}$. All the results match the analytic forms evaluated in Eqs. \eqref{dvz} amd \eqref{cvz}. These results demonstrates that the non-Markovian dynamics facilitated by the bound state can restore quantum superiority for dense coding under environmental noise.

Figure \ref{s} explores the combined effects of the environmental spectral index $s$ and the coupling constant $\eta$. Figure \ref{s}(a) displays the long-time limit of $|u(t)|$ in the $s$-$\eta$ plane, with the same threshold lines and bound-state criterion. The energy spectrum versus $s$ at $\eta=0.01$ in Fig. \ref{s}(b) confirms that the bound state exist for $s<1.0$ or $s>2.0$, where $|u(t)|$ tends to a finite value consistent with $Z$ under the analytical condition $\gamma(s)\geq \omega_0/\eta\omega_c$. The corresponding $C(t)$ is presented in Fig. \ref{s}(c) and Fig. \ref{s}(d) for the discrete- and continuous-variable systems, respectively. Once the bound state is present, $C(t)$ clearly approaches a finite asymptotic value matching the analytical result (red curves).  For a detailed comparison, Fig. \ref{s}(e) and Fig. \ref{s}(f) examine the exact dynamics at $s=2.2$. Consistent with Fig. \ref{eta}, the discrete-variable capacity converges to a steady value yielding a quantum advantage, while the oscillatory continuous-variable capacity retains its quantum advantage within the long-time envelope. The same conclusion can be drawn by jointly changing $s$ and $\omega_c$, see Fig. \ref{n}. All the above results confirm that the quantum advantage of dense coding can be sustained under environmental noise in the presence of a bound state. They reveal the clear advantage of the non-Markovian effect and directly demonstrate the pivotal role of the bound state.

\section{Discussion and conclusion}\label{discussion}
In summary, we have proposed a mechanism to sustain quantum advantage in both discrete- and continuous-variable quantum dense coding under environmental noise. By investigating the non-Markovian decoherence dynamics, we find that the performance of dense coding is sensitively determined by the energy spectrum of the total system formed by the channel subsystems and their environments. When a bound state emerges, the long-time capacity of both dense coding schemes exceeds their respective classical limits, thereby preserving quantum advantage. This performance is unattainable in previous studies \cite{PhysRevA.85.052306,PhysRevA.90.022319,Shadman_2010,article,PhysRevApplied.13.034029,PhysRevApplied.19.064015,PhysRevLett.126.250501,PhysRevA.84.042309,e24081114,PhysRevA.93.032135,Liu_2016,PhysRevA.109.032616,Zhao2019,PhysRevA.82.032340,islam_optimizing_2021,AIACHE2024129320}. The key physical mechanism underpinning this sustained advantage is information backflow, enabled by the structured system-environment coupling. Unlike phenomenological approaches that simulate non-Markovianity using static Kraus operators \cite{PhysRevA.109.032616}, our mechanism is dynamically generated and thus intrinsically capable of sustaining quantum advantage over extended timescales. The ability to sustain quantum advantage under practical noise conditions has significant implications. It offers a passive protection strategy via environmental engineering, which represents a fundamental change from fighting noise to harnessing it. The long-lived quantum advantage enables practical applications such as high-capacity quantum communication \cite{ren_deterministic_2026} and can be integrated into quantum networks. Moreover, while recent breakthroughs in quantum error correction have shown that surpassing the break even point is critical for practical quantum advantages \cite{brock_quantum_2025}, our mechanism provides an alternative pathway to achieve this milestone through environmental engineering. Therefore, our findings lay a foundation for noise-resilient quantum communication protocols and offer a feasible strategy for high-capacity quantum communication in practical settings.
	
\begin{acknowledgments}
The work is supported by the National Natural Science Foundation of China (Grant No. 92576202), the Quantum Science and Technology-National Science and Technology Major Project (Grant No. 2023ZD0300904), the Natural Science Foundation of Gansu Province (Grants No. 26RCKA011 and No. 25JRRA799), the Fundamental Research Funds for the Central Universities (Grant No. lzujbky-2026-jdzx02), and the ‘111 Center’ (Grant No. B20063). 
\end{acknowledgments}
	
\bibliography{codingnew}	
\end{document}